\documentclass[12pt]{iopart}
\usepackage{amssymb}
\usepackage[english]{babel}
\usepackage{float}
\usepackage{wrapfig}
\usepackage{svg}
\usepackage{graphicx}
\usepackage{dcolumn}
\usepackage{bm}
\usepackage{hyperref}
\hypersetup{
    colorlinks=true,
    linkcolor=blue,    
    citecolor=blue,    
    urlcolor=blue      
}

\begin{document}
\title[Empirical scaling laws for self-focused laser pulses in nitrogen plasmas]{Empirical scaling laws for self-focused laser pulses in nitrogen plasmas}
\author{L.~Martelli$^{1,2}$, I.~A.~Andriyash$^1$, J.~Wheeler$^1$, H.~Kraft$^2$ X.~Q.~Dinh$^2$ and C.~Thaury$^1$}
\address{$^1$ Laboratoire d’Optique Appliqu\'ee, ENSTA Paris, CNRS, \'Ecole Polytechnique,\\Institut Polytechnique de Paris, 828 Bd des Mar\'echaux, 91762 Palaiseau, France}
\address{$^2$ THALES AVS - MIS, 2 Rue Marcel Dassault, Vélizy-Villacoublay, 78140, France}
\ead{cedric.thaury@ensta.fr}

\begin{abstract}
We investigate the interaction between a superintense laser pulse and a nitrogen plasma with densities exceeding $10^{19}\,$cm$^{-3}$, using particle-in-cell simulations. Such configurations have recently demonstrated the capability to produce highly charged electron beams (i.e., $>10\,$nC) with 1\,J-class lasers, a significant step toward high-average-current laser-plasma accelerators. Our study focuses on analyzing the impact of laser self-focusing on laser dynamics, leading to scaling laws that characterize beam diffraction, wakefield amplitude and plasma structures, providing important insights of this interaction regime.
\end{abstract}


\section{Introduction}
Laser-plasma accelerators (LPAs)~\cite{Tajima} enable accelerating gradients that are three to four orders of magnitude higher than those of conventional systems (e.g., $\sim$100\,MV/m), paving the way for a new generation of compact, cost-effective accelerators. Thanks to their ultrashort bunch durations, LPAs can also produce kA-level peak currents~\cite{Lundh, Couperus, Kurz, Winkler}, making them promising candidates for applications such as ultrafast imaging~\cite{Wood2018} and ultra-high-dose-rate radiotherapy~\cite{Vozenin2022, Flacco2024}. However, due to the low repetition rate of current high-intensity laser systems, LPAs presently achieve average currents of at most tens of nA~\cite{Couperus, Rovige2020}—over three orders of magnitude lower than those produced by conventional accelerators. Increasing the average current is a crucial step toward enabling laser-plasma-based applications, such as the Multiscan 3D project~\cite{multiscan, Martelli_thesis}, which aims to develop the first LPA-driven 3D Bremsstrahlung X-ray tomography system for border inspection. One promising approach to boost the average current is to directly increase the charge per electron bunch. The use of high-Z gases such as nitrogen with plasma densities $n_e>10^{19}\,$cm$^{-3}$ has recently been proven effective in increasing the beam charge above $10\,$nC~\cite{JFeng, Martelli}. This, along with recent advancements in laser technology that promise the development of $100\,$W (i.e., 1\,J/shot at 100\,Hz) Ti:Sapphire laser systems~\cite{Pellegrina2022, Kiani2023}, potentially paves the way for $\mu$A-level LPAs.

In this paper, we investigate a rarely explored interaction regime between a superintense (i.e., $I>10^{18}\,$W/cm$^{-2}$) laser pulse and a nitrogen plasma with a density $n_e\geq0.02\,n_c$, where $n_c(\mbox{cm$^{-3}$}) = 1.1\times10^{27}/\lambda_0^2(\mbox{nm})$ is the critical plasma density, and here $\lambda_0=800\,$nm is the laser central wavelength. Using particle-in-cell (PIC) simulations, we study the effect of self-focusing on the laser beam evolution in plasma. This, in turn, allows the retrieval of empirical scaling laws on relevant quantities such as the maximum laser normalized vector potential in plasma $a_{P,max}$ and the depletion length $L_{pd}$. From our study, we also notice that under the configurations of interest, the laser creates a plasma channel through its ponderomotive force, as observed in previous works~\cite{Cohen2024}.

\section{Methods}
We perform the numerical study using the 3D code Fourier-Bessel particle-in-cell (FBPIC), which employs a cylindrical grid with azimuthal decomposition~\cite{fbpic}. Our simulation setup includes a linear density ramp extending over $30\,\mu$m, allowing the laser pulse to focus with minimal energy losses over a $1500\,\mu$m plateau.

The nitrogen gas is preionized up to N$^{+3}$, corresponding to the first three L-shell electrons for computational ease. Under the conditions of interest, the laser pulse can further ionize the gas to at least N$^{+5}$ with ease. Thus, we consider four different plateau plasma densities corresponding to the full L-shell ionization (i.e., N$^{+5}$), namely between $n_e=0.02\,n_c$ and $n_e=0.18\,n_c$. The laser is assumed as a $30\,$fs-Gaussian beam propagating along $z$ and polarized in the $x$ direction, with energy $E_L$ ranging from $0.05\,$J to $1\,$J, and the laser waist is $w_0=3\,\mu$m. Regarding the numerical parameters, we employ a $(r,z)$ mesh with $\Delta z = \lambda_0 / 24$ and $\Delta r = 5\,\Delta z$, where $\lambda_0 = 800$\,nm is the laser wavelength. Lastly, we consider three azimuthal modes $(m=0-2)$ and the number of macroparticles per cell along $r,\,z$ and $\theta$ is 1, 1 and 4 respectively.

\section{Numerical results}
\begin{figure}[htbp]
    \centering
    \includegraphics[width=30pc]{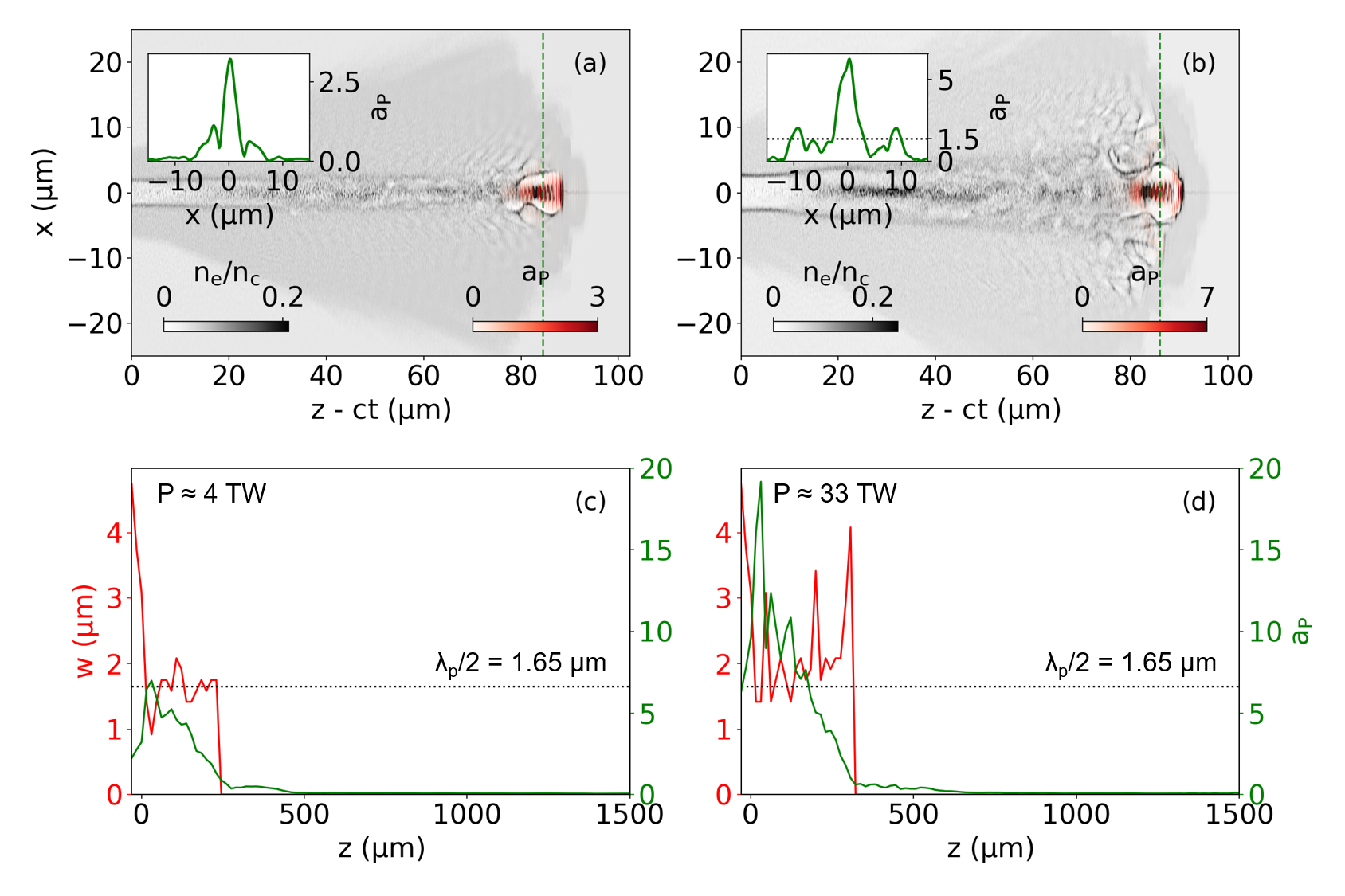}
    \caption{FBPIC results for $n_e=0.06\,n_c$. The gray colormap represents density snapshots at (a) $E_L=0.12$\,J and at (b) $E_L=1$\,J. The red colormap represents the laser envelope, expressed in terms of the normalized vector potential in plasma $a_P$. The insets refer to the laser transverse profile along the green dashed line. (c-d) Beam waist and normalized vector potential in plasma for $E_L=0.12$\,J and $E_L=1$\,J.}
    \label{fig:diff_laser}
\end{figure}

We first discuss laser diffraction considering two simulations at $n_e=0.06\,n_c$ with laser energies $E_L=0.12\,$J and $E_L=1\,$J shown in Fig.~\ref{fig:diff_laser}(a) and (b) respectively. Specifically, the gray colormap illustrates the plasma density $n_e/n_c$, while the red colormap represents the laser, expressed in terms of the normalized vector potential $a_P$ in the co-moving box reference frame. On each snapshot, we also present the transverse laser profile along the green dashed line. From these images, we can already observe the emergence of complex plasma structures that we will discuss in the following. Moreover, we note that the laser stops having a Gaussian shape, as highlighted by the appearance of sidelobes due to beam-break up~\cite{AThomas}, a phenomenon seeded by self-focusing. These sidelobes can reach sufficiently high intensities to even cause K-shell ionization~\cite{ADK} (i.e., for $a_P\gtrsim1.5$), as shown in the inset of Fig.~\ref{fig:diff_laser}(b), where the black dotted line corresponds to $a_P=1.5$. The density colormap of Fig.~\ref{fig:diff_laser}(b) also highlights the excitation of plasma modes far from the laser axis (i.e., at a distance exceeding $\sim2 w_0$), due to the presence of these intense sidelobes.

In Fig.~\ref{fig:diff_laser}(c) and (d), we plot the laser spot size $w$ and normalized vector potential over the depth in plasma $z$, for the configurations presented in Fig.~\ref{fig:diff_laser}(a) and (b) respectively. Considering the range of laser energy and plasma density of interest, it is straightforward to notice that the laser power exceeds the critical power (i.e., $P>P_c\mbox{(GW)} \approx 17 n_c/n_e\,$), which ensures the occurrence of self-focusing~\cite{Esarey2009}. Indeed, for $n_e=0.06\,n_c$ we estimate $P_c\approx0.3\,$TW, while the laser power is $P=4\,$TW and $P=33\,$TW for $E_L=0.12\,$J and $E_L=1\,$J respectively. Hence, from Fig.~\ref{fig:diff_laser}(c) and (d) we deduce that self-focusing maintains the laser focused in plasma for distances several times the Rayleigh length ($Z_R \approx 35\,\mu$m). For instance, Fig.~\ref{fig:diff_laser}(c) shows that the laser remains guided over a distance $L \approx 7\,Z_R$ before being depleted. Moreover, we notice that the laser waist oscillates around half of the plasma wavelength $\lambda_p/2$, as already observed in other works~\cite{AThomas, SMangles}. This, in turn, means that higher plasma densities lead to a tighter focusing and, consequently, to higher values of the normalized vector potential. Indeed, as it is possible to infer from Fig.~\ref{fig:diff_laser}(c) and (d), the normalized vector potential in plasma can reach a maximum up to $a_{P,max}\approx6$ and $a_{P,max}\approx20$ shortly after the density ramp with a laser energy of $E_L=0.12$\,J and $E_L=1$\,J respectively. In a vacuum, instead, we recall that with $E_L=0.12$\,J the maximum normalized vector potential is $a_0 = 3.6$, while with $E_L=1$\,J we get $a_0 = 10$.

\begin{figure}[htbp]
    \centering
    \includegraphics[width=20pc]{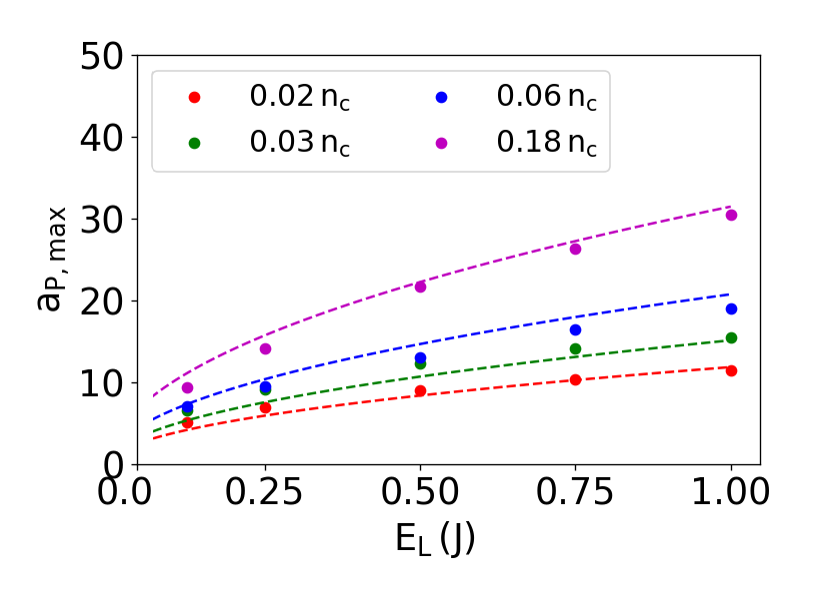}
    \caption{Maximum normalized vector potential in plasma for different plasma densities and laser energies. The dots refer to the values obtained through FBPIC simulations, while the discontinuous lines have been obtained empirically.}
    \label{fig:a0_mod}
\end{figure}

The colored dots in Fig.~\ref{fig:a0_mod} represent the values of $a_{P, max}$ obtained with FBPIC for different plasma densities and laser energies. Here, we notice an increase in the maximum normalized vector potential in plasma with the plasma density. As already mentioned, this effect arises from self-focusing, which reduces the laser waist to approximately $\lambda_p/2$ and hence the laser intensity varies as $1/\lambda^2_p\propto n_e$.
From the numerical study, we are also able to deduce the following empirical law for the maximum normalized vector potential in plasma:
\begin{eqnarray}
    a_{P, max} \approx 91\,\sqrt{n_e/n_c \times E_L(\mbox{J})}\,(1-n_e/n_c),
    \label{eqn:a0_mod}
\end{eqnarray}
where the factor $\sqrt{n_e/n_c}$ accounts for the reduced beam waist due to self-focusing and $(1-n_e/n_c)$ for other phenomenons such as laser depletion, reflection and dispersion. The dashed lines in Fig.~\ref{fig:a0_mod} have been obtained with Eq.~(\ref{eqn:a0_mod}), proving the good agreement with the numerical results.

\begin{figure}[htbp]
    \centering
    \includegraphics[width=20pc]{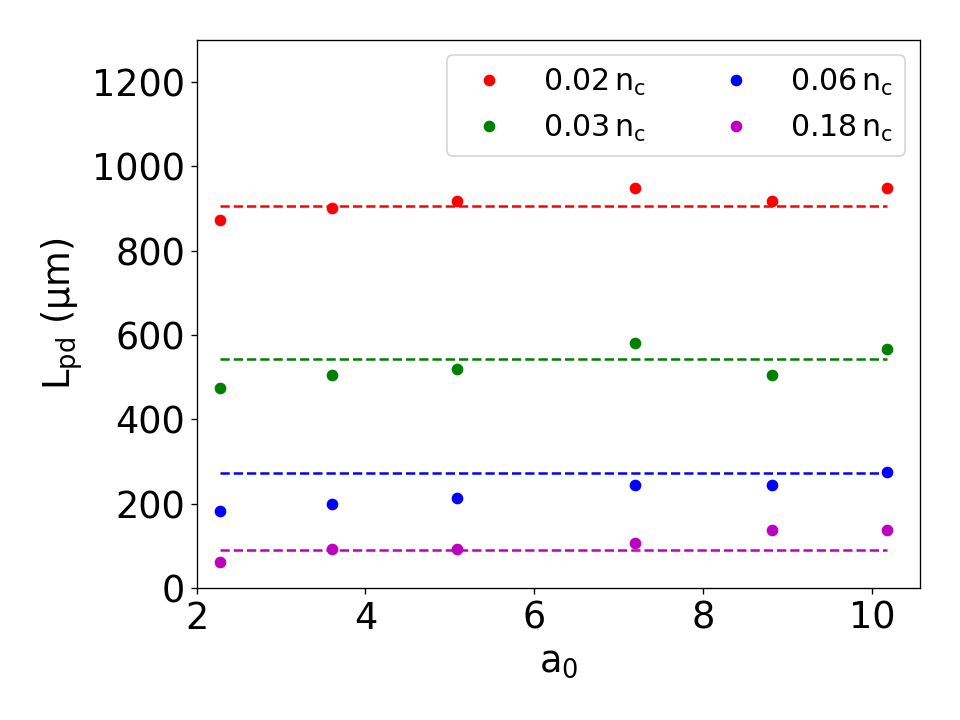}
    \caption{Pump depletion length for different densities, across different $a_0$'s. The dots represent the numerical results, while the dashed lines are a fit of the form $L_{pd}(\mu\mbox{m}) \approx 16\,n_c/n_e$.}
    \label{fig:Lpd}
\end{figure}

We now intend to examine the pump depletion length $L_{pd}$. Here, we define $ L_{pd} $ as the distance in plasma where the laser remains such that $ a_P > 1 $, namely, ensuring the generation of a non-linear wakefield. This quantity is relevant as it describes one of the main mechanisms that can hinder the electron energy gain in laser-plasma accelerators and potentially limit the maximum achievable charge. Hence, in Fig.~\ref{fig:Lpd} we present the pump depletion length (dots) for different densities across the vacuum maximum normalized vector potential $a_0$. In this figure, we observe that the pump depletion length displays a small linear increase over $a_0$, while it strongly depends on the plasma density. Specifically, we deduce the following empiric scaling of the pump depletion length
\begin{eqnarray}
    L_{pd} \propto \frac{a_0 + K}{n_e}\,n_c,
    \label{eqn:lpd}
\end{eqnarray}
where $K$ is a constant we numerically determined and since $K\gg a_0$ for $a_0\in\left[2,10\right]$, we can write $L_{pd}\propto \,n_c/n_e$, which leads in practical units to
\begin{eqnarray}
    L_{pd}(\mu\mbox{m}) \approx 16\,n_c/n_e.
\end{eqnarray}
A similar scaling was obtained by Lu~\cite{Lu2007} for a matched beam with $k_p w_0 = 2\sqrt{a_0}$ and $a_0>2$, while Gordienko~\cite{Gordienko2005} estimated $L_{pd}\propto a_0\,\,n_c/n_e$, assuming a circularly polarized laser and $k_p w_0 = 1.12\sqrt{a_0}$. Here, $k_p=2\pi/\lambda_p$ is the plasma wavenumber.

\begin{figure}[!htb]
    \centering
    \includegraphics[width=28pc]{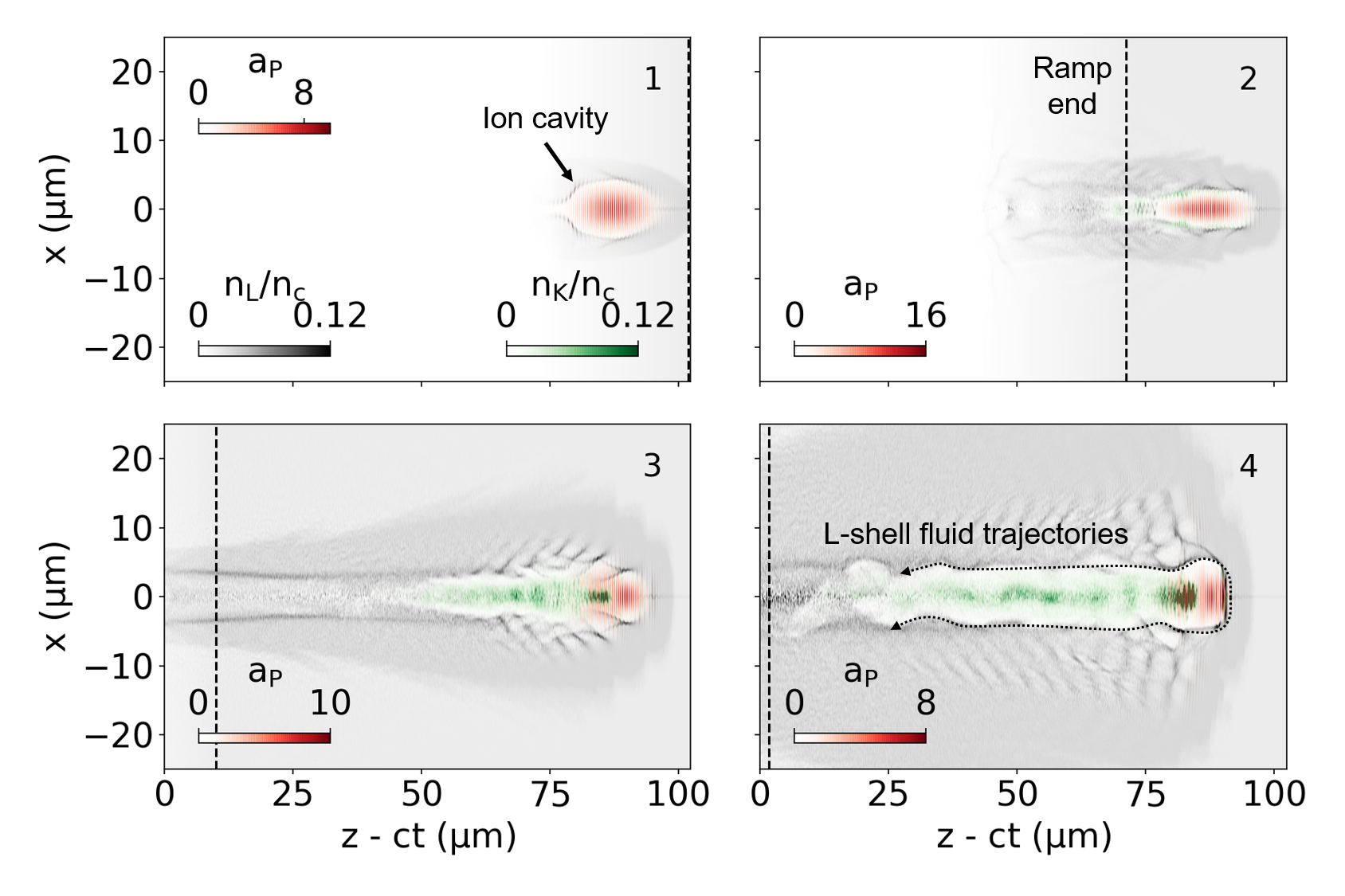}
    \caption{$(1-4)$ Plasma density snapshots of four consecutive iterations. The gray and green colormaps refer to the L- and K-shell electron density. In red we depict $a_P$. These panels allow to observe the generation of a channel-like structure.}
    \label{fig:channel_form}
\end{figure}

Concerning the structures in plasma, the numerical analysis underlines the formation of a long and rapidly changing channel-like structure shortly after the end of the density ramp. Fig.~\ref{fig:channel_form} depicts the density snapshots of four consecutive iterations $(1-4)$, where the gray (green) colormap refers to L-shell (K-shell) electrons, while the red colormap is $a_P$. Here, we notice that the laser's ponderomotive force radially pushes L-shell electrons upon interacting with the density plateau, creating a copropagating ion cavity $(1)$. Simultaneously, a significant number of K-shell electrons are ionized into the cavity. The continuous loading leads to the accumulation of K-shell electrons on the laser axis $(2-3)$. Thus, due to the high concentration of K-shell electrons on axis, L-shell electrons cannot close their trajectories to form ion cavities. Instead, they slip to the rear, forming a channel-like sheath, surrounding K-shell electrons $(4)$.

\begin{figure}[htbp]
    \centering
    \includegraphics[width=28pc]{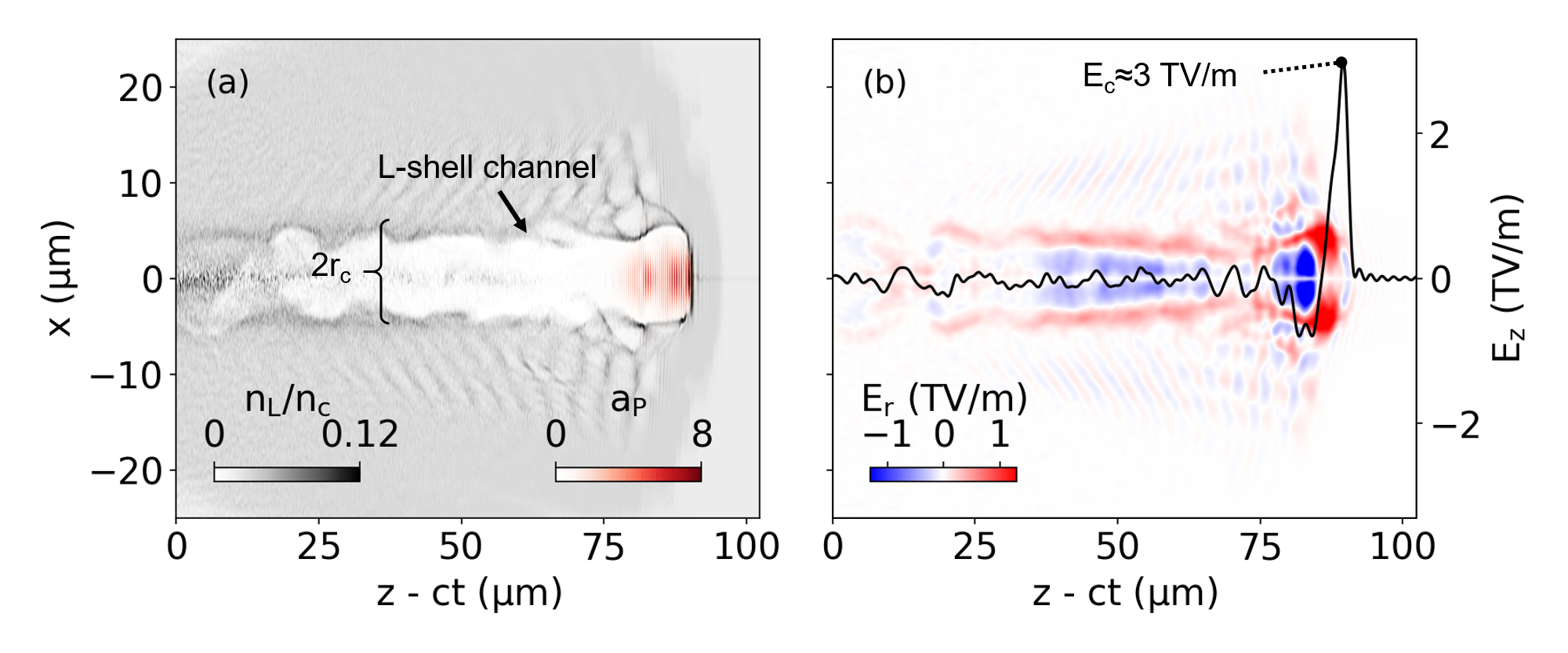}
    \caption{(a) Plasma density snapshot of the last panel of Fig.~\ref{fig:channel_form}. The gray colormap refers to L-shell electrons, while the red one refers to the laser. (b) Radial wakefield map ($E_r$) and longitudinal wakefield ($E_z$) corresponding to the snapshot shown in (a).}
    \label{fig:channel}
\end{figure}

In Fig.~\ref{fig:channel}(a) we plot the last density snapshot of Fig.~\ref{fig:channel_form}. For visual purposes, in Fig.~\ref{fig:channel}(a) we only display the density of L-shell electrons, highlighting the channel structure. We estimate a maximum channel length of $\sim70\,\mu$m, corresponding to $\sim 15\lambda_p$ and a radius of around $r_c=6\,\mu$m, exceeding the ion cavity radius in the blowout regime~\cite{Lu2007}, namely $r_b = 2\sqrt{a_0}/k_p\approx4.7\mu$m. Compared to Ref.~\cite{Lu2007}, here, the laser is experiencing a much stronger self-focusing, reaching higher intensities in plasma than in a vacuum, as previously discussed.
Thus, replacing $a_0$ with Eq.~(\ref{eqn:a0_mod}) in $r_b = 2\sqrt{a_0}/k_p$ we find that the channel's maximum radius is in practical units
\begin{eqnarray}
    r_c(\mu\mbox{m}) \approx 2.4\,E^{1/4}_L(\mbox{J})\,(n_e/n_c)^{-1/4}\,(1 - n_e/n_c)^{1/2}.
    \label{eqn:rad_mod}
\end{eqnarray}
With Eq.~\ref{eqn:rad_mod} we now yield a radius of $\sim 5.7\,\mu$m, which is in better agreement with the simulation results, as shown in Fig.~\ref{fig:channel}(a).

Fig.~\ref{fig:channel}(b), instead, illustrates the radial wakefield map ($E_r$) and the longitudinal wakefield ($E_z$) corresponding to the iteration of Fig.~\ref{fig:channel}(a). At the front of the channel structure, the laser ponderomotive force allows for an effective charge separation, generating an ion cavity. Here, the wakefield reaches a maximum of $\sim 3$\,TV/m, exceeding the value predicted by Lu~\cite{Lu2007} with
\begin{eqnarray}
    E_0 (\mbox{GV.m$^{-1}$}) \approx 96 \sqrt{a_0 \, n_e (10^{18}\mbox{cm$^{-3}$})},
    \label{eqn:dawson_lu}
\end{eqnarray}
namely $E_0 \approx 2.2$\,TV/m.

As in our prior discussion, we replace $a_0$ with $a_{P,max}$ in Eq.~(\ref{eqn:dawson_lu}) and we find the following expression for the maximum wakefield
\begin{eqnarray}
    E_c(\mbox{GV.m$^{-1}$}) \approx 4.1\times10^{4}\,E^{1/4}_L(\mbox{J})\,(n_e/n_c)^{3/4}\times(1-n_e/n_c)^{1/2}.
    \label{eqn:wake_mod}
\end{eqnarray}
Substituting $E_L=1\,$J and $n_e/n_c = 0.03$ in Eq.~(\ref{eqn:wake_mod}) yields $E_c = 2.8\,$TV/m, which is close to the value obtained with the simulation. Lastly, in Fig.~\ref{fig:channel}(b) we notice that the longitudinal wakefield quickly drops to a maximum of around $0.2\,$TV/m inside the channel, due to the screening effect of K-shell electrons. Moreover, the massive accumulation of K-shell electrons on the axis results in a defocusing radial wakefield, which significantly differs from that of a classical plasma cavity.

\begin{figure}[!htb]
    \centering
    \includegraphics[width=28pc]{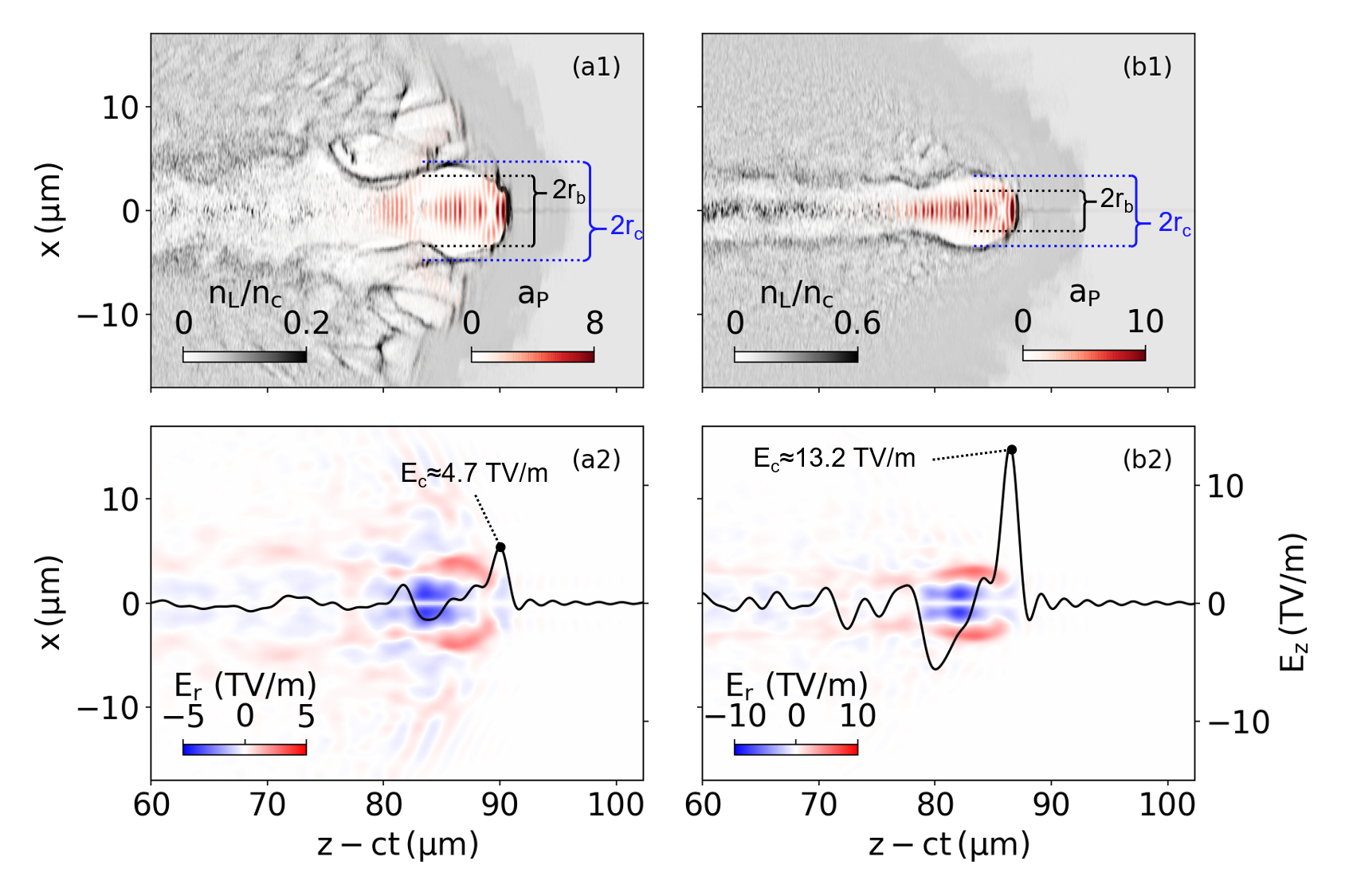}
    \caption{Results with $E_L=1\,$J and (a1-2) $n_e = 0.06\,n_c$ and (b1-2) $n_e = 0.18\,n_c$. (a2, b2) Radial wakefield in blue and red, while the black curve refers to the longitudinal wakefield.}
    \label{fig:rad_wake}
\end{figure}

To further validate the scaling in Eq.~(\ref{eqn:rad_mod}) and (\ref{eqn:wake_mod}), in Fig.~\ref{fig:rad_wake} we present the results for $E_L=1\,$J with $n_e = 0.06\,n_c$ (first column) and $n_e = 0.18\,n_c$ (second column). The top row shows L-shell density snapshots at the iterations where we reach the maximum wakefield. Here, it is possible to notice that Eq.~(\ref{eqn:rad_mod}) provides a good estimate of the maximum channel radius. In the bottom row, instead, we illustrate the radial and longitudinal wakefield, which reaches a maximum of around $4.7\,$TV/m and $13.2\,$TV/m at $n_e = 0.06\,n_c$ and $n_e = 0.18\,n_c$ respectively. In the first case, Eq.~(\ref{eqn:wake_mod}) yields $4.6\,$TV/m, while in the second $10\,$TV/m, proving the good agreement with the numerical results.

\section{Conclusion}
In this paper, we conduct a detailed numerical campaign to investigate laser diffraction in nitrogen plasmas with densities exceeding $n_e=0.02\,n_c$. Such conditions prove capable of delivering few-MeVs electron beams with charges surpassing $10\,$nC--an important milestone toward $\mu$A laser-plasma accelerators~\cite{JFeng, Martelli}.

Notably, we observe the occurrence of beam-breakup initiated by self-focusing, coherent with prior works in the scientific literature~\cite{AThomas}. Furthermore, we demonstrate that the self-focused laser reaches a beam waist of approximately $\lambda_p/2$~\cite{AThomas, SMangles}, which is smaller than the vacuum focus and leads to a significant enhancement of the laser normalized vector potential, up to three times the value in vacuum. Based on these simulations, we derive empirical scaling laws for both the vector potential in plasma and the corresponding laser depletion length.

Additionally, our study highlights the formation of a massive channel-like structure~\cite{Cohen2024}, due to a high concentration of K-shell electrons on axis. From the analysis of such structures, we determine scaling laws for their characteristic radius and peak wakefield amplitude.

\section*{Acknowledgments}
This project has received funding from the European Union’s Horizon 2020 research and innovation program under grant agreement n°101020100.

\section*{References}
\bibliography{biblio}
\end{document}